\documentclass[10 pt,onecolumn,draft]{article}
\usepackage{amsmath}
\usepackage{graphicx}
\usepackage{amsfonts}
\usepackage{amssymb}

\begin{document}

\title{Quantum Message Disruption: A Two-State Model}
\author{Paul B. Kantor\thanks{Rutgers University. kantor@scils.rutgers.edu.  APLab Technical Report APLab/TR-00/3}}
\maketitle
\begin{abstract}
A game in which one player makes unitary transformations of a simple system,
and another seeks to confound the resulting state by a randomly chosen action
is analyzed carefully. \ It is shown that the second player can reduce any
system to a completely random one by rotation through an angle of 120 degrees,
about an axis chosen at random. If, on the other hand, the second player is
forced to behave ``classically'' by reducing the wave function, then the first
play retains an advantage, which the second player may eliminate by repeated
measurement using randomly selected bases.
\end{abstract}

%
%

\onecolumn

\makeatletter
\global\@specialpagefalse
\let\@evenhead\@oddhead
%
%
%

\let\@evenfoot\@oddfoot
\makeatother
Meyer [1] discusses a problem involving systems with two states, characterized
informally as a ``coin toss'', played between two individuals one described as
Quantum Mechanical (Q) and the other as Classical (P). The object of the
''game'' is to predict the final state of the system. His analysis shows that
the Quantum Mechanical player, if permitted to play first and to play third
can always beat the Classical player who makes only the second move. This is
in contrast to the classical situation, where the middle player may flip the
coin at random to achieve a fair outcome. The analysis seems incomplete, since
the clear advantage to the first player implies that the second player, try as
he may, cannot leave the coin in a state unknown to the first player.
Formally, the second player in [1] applies, with probability 0.5, a known
unitary operator to the state produced by the first player. Although this
operator can be interpreted as ``tossing the coin'', (that is, it interchanges
the two possible initial states), it is not the only operator that can do so.
Specifically, using a natural notation, it represents a rotation of 180
degrees about the x-axis.

The same effect could be achieved by rotation of 180 degrees about the y-axis,
or in fact about any axis perpendicular to the z-axis. Since P knows that Q
has done something to the ``coin'', it is not clear that this transformation
will flip the coin. The analysis in [1] uses Q's knowledge of P's strategy and
has Q rotate the original state so that the original quantization axis lies
along the x-axis, ensuring that the second player has no effect whatsoever on
the state of the system. In short, player Q has an unfair advantage because he
knows exactly what P will do. \ More realistically, P either is bound under
the rules of the game to use the specific rotation given in [1], or he can
confound Q by choosing a different rotation, based on his knowledge of Q's
strategy. \ For example, he might choose to rotate the system 180 degrees
about one of two axes chosen at random, from among two mutually perpendicular
axes orthogonal to the original polarization direction. \ We will not pursue
this approach further because it assumes that both players have agreed on the
orientation of a coordinate system in space when, in fact, they need not do so
and could intentionally lie in order to gain advantage.

We present here a more detailed analysis of the situation described in [1] and
find that the middle player, P, by following a genuinely mixed strategy [von
Neumann and Morganstern; 2] can in fact achieve parity. The heart of the
analysis is to understand what it means to say that the second player is
``classical''. \ In the interpretation of [1] it was taken to mean that P has
available only two unitary transformation to be applied to the state of the
system, and chooses a mixture of those two. \ \ The resulting state of the
system is a mixed state and will, in general, have non-zero entropy, that is,
it will not correspond to a determinate value of the original observable.
\ Since entropy is preserved by unitary transformations, it would follow that
Q cannot bring it to a determinate state. \ In [1] this difficulty is avoided
by making Q unfairly aware of the transformation to be used by P. \ Q first
brings the system to an eigenstate of the operator to be used by P. \ The
transformation applied by P does not change this state, entropy is not
increased, so Q knows the state after P has played, and \ of course wins.

There are two ways that P can avoid this embarrassment. \ One is to choose his
rotation of the system at random, rather than choosing the one that Q expects.
\ We will show that this strategy, and the rather surprising choice of a
rotation of 120 degrees, makes the game fair. \ The other course is to
``behave classically'', that is, to make a measurement of the system, reducing
it to an eigenstate of the measurement operator. This will increase the
entropy of the system, ensuring that Q cannot restore it to a pure state.
\ However we find in this case (again surprisingly) that even if the
measurement is chosen at random, Q retains an advantage. \ P can make this
advantage arbitrarily small by repeated (random) measurement of the system, in
a reversal of the usual encoding processes, where it is the probability of
error, rather than of advantage, that is reduced.

To summarize the argument of [1] very briefly: player P will use the strategy
$S[1]=(p,(1-p)F)$ which transforms a state described by the density matrix
$\rho$ into the mixture \ $p\rho+(1-p)F\rho F^{\dagger}.$ \ Hence, if the
matrix $\rho$ \ commutes with $F$ \ then the entire mixture is simply equal to
$\rho.$ \ In [1] player Q takes advantage of his knowledge of the operator $F$
to transform the original state into a particular eigenstate of $F.$ \ Of
course if player P has \emph{read} reference [1] then he will choose a
different operator $F,$ and foil this scheme. \ 

To formalize our analysis, consider the simplest quantum mechanical example of
a two-state system, which may be thought of as representing a particle of spin
one-half. We continue to allow player Q first and third plays. Technically, in
changing from the loose natural language of coin toss to the formal language
of spin one-half systems, we are not only following the path suggested in
Ref.[1], but also asserting the fundamental point that the number of degrees
of freedom of a system in the real world is a matter of fact, and not a matter
of how we choose to discuss it. A classical coin has many of degrees of
freedom (six classical degrees as a rigid body, plus internal degrees of
freedom, admitting an enormous number of physical states) and is not the
appropriate model for the game discussed in [1].

We suppose that Q, following [1] has taken the system which starts in a pure
state (see von Neumann, [3]), and has rotated it by an unknown amount about an
unknown axis. It remains in a pure state. The problem for the second player is
to move the system into a ``totally mixed'' state. This can not be done (the
fact that was exploited in [1]) by the application of a unitary
transformation, or even a mixed strategy built on only two such
transformations. It can however be done by the application of a ``randomly
selected'' unitary transformation, which throws the system into a state
represented by a density matrix that is the linear superposition of all the
transformations of the original density matrix.

The goal of P is to reduce the system to a fully unpolarized state, whose
density matrix is a multiple of the identity. This characterization is
invariant under rotations, so if P can achieve it for a particular $\rho$, it
does not matter that he does not know the orientation of the axis used by the
first player. We exploit this fact by assuming that when P begins his play the
system is in the pure state characterized by the density matrix $\rho_{1}.$%
\begin{equation}
\rho_{1}=\left\|
\begin{array}
[c]{cc}%
1 & 0\\
0 & 0
\end{array}
\right\|
\end{equation}

P's problem is to find a mixture of transformations that brings this to a
multiple of the identity. This problem is easily solved using a standard
representation of rotation in terms of the Pauli Spin Matrices. We conjecture,
on purely physical grounds, that the solution will be given by a rotation
through some fixed angle $\theta$ about an arbitrarily chosen axis
$\widehat{n}$. The result can be represented by the integral given in Equation
\ref{define_state2}.
\begin{equation}
\rho_{2}=%
{\displaystyle\int}
d\widehat{n}U(\widehat{n},\theta)\rho_{1}U(\widehat{n},\theta)^{\dagger}
\label{define_state2}%
\end{equation}

Using the standard form for the rotation in two-dimensional state space, we
have:
\begin{equation}%
\begin{array}
[c]{c}%
U(\widehat{n},\theta)=e^{i\theta\sigma\cdot\widehat{n}/2}\\
=\cos\theta/2+i\sin\theta/2\sigma\cdot\widehat{n}\\
=\left\|
\begin{array}
[c]{cc}%
\cos\theta/2+in_{z}\sin\theta/2 & i(n_{x}-in_{y})\sin\theta/2\\
i(n_{x}+in_{y})\sin\theta/2 & \cos\theta/2-in_{z}\sin\theta/2
\end{array}
\right\|
\end{array}
\end{equation}

Hence the integral becomes (we write $c=\cos\theta/2;s=\sin\theta/2)$%

\[
\rho_{2}=%
{\displaystyle\int}
d\widehat{n}U(\widehat{n},\theta)\rho_{1}U(\widehat{n},\theta)^{\dagger}%
\]

\bigskip$=%
{\displaystyle\int}
d\widehat{n}\left\|
\begin{array}
[c]{cc}%
c^{2}+n_{z}^{2}s^{2} & (c-in_{z}s)(in_{x}+n_{y})s\\
CompConj & (n_{x}^{2}+n_{y}^{2})s^{2}%
\end{array}
\right\|  $

Here $CompConj$ represents the complex conjugate of the off-diagonal matrix
element shown. Note that $\theta$ is not a variable of integration here, but a
parameter to be determined. Since $n_{x}^{2}+n_{y}^{2}+n_{z}^{2}=1$, and the
integration is spherically symmetric, $\int d\widehat{n}\,\,n_{x}n_{y}=0$,
while $\int d\widehat{n}\,\,n_{x}^{2}=\int d\widehat{n}\,\,n_{y}^{2}=\int
d\widehat{n}\,\,n_{z}^{2}=1/3$, etc. Hence:
\begin{equation}
\rho_{2}=\left\|
\begin{array}
[c]{cc}%
\cos^{2}\theta/2+(1/3)\sin^{2}\theta/2 & 0\\
0 & 2/3\sin^{2}\theta/2
\end{array}
\right\|
\end{equation}

This is a multiple of the unit matrix only if cos$^{2}\theta/2=1/3\sin
^{2}\theta/2,$ or cos$^{2}\theta/2=1/4.$ Thus $\theta/2=60^{\circ},$ and
$\theta=120^{\circ}.$\ \ \ Thus, if P's strategy is to rotate the system by
$120^{\circ}$about an axis chosen at random, the resulting state will be fully
mixed and will remain so despite the best efforts of the first player to
reorder it.

It seems puzzling that the required angle is more than $90^{\circ}$. I believe
that this is due to the fact that by accepting rotation about all axes, we
have allowed for rotations that are ``nearly aligned'' with the axis chosen at
random by the first player. But those rotations have relatively little effect
on the state of the system, and so all rotations must be through an angle
greater than 90$^{\circ}$ in order to achieve the required full mixing of the
original state. The specification of the state when P begins has no effect on
these conclusions. If, instead, the state has been rotated, that rotation can
be removed from the rotations used by P, through a change of the variable of integration.

Next consider the situation in which P behaves classically: that is, performs
a measurement on the system. \ Clearly if P knows that the system is fully
polarized along the axis $\widehat{n}$ \ he can mix the system by making a
measurement along any axis orthogonal to $\widehat{n}.$ \ However, either
because of the lack of communication between the players, or because Q wishes
to prevent this, the orientation of the system at the end of the first play
should be unknown to P. \ Hence his action is, in effect, a measurement along
an axis chosen at random with respect to the state of the system when he
receives it.

The result of a measurement corresponding to the basis states $|\beta>$ \ is
to transform the density matrix from $\rho$ \ to the mixture:
\begin{equation}
\rho\rightarrow\sum_{\beta}|\beta><\beta|\rho|\beta><\beta|
\end{equation}

The eigenstates of spin 1/2 in a \ direction $\widehat{n}(\theta,\phi)$ \ are
given by (we write $z=\cos\theta$):
\begin{align}
\beta_{+}  &  =\frac{1}{\sqrt{2(1+z)}}\left\|
\begin{array}
[c]{c}%
1+n_{z}\\
n_{x}+in_{y}%
\end{array}
\right\| \\
\beta_{-}  &  =\frac{1}{\sqrt{2(1-z)}}\left\|
\begin{array}
[c]{c}%
1-n_{z}\\
-(n_{x}-in_{y})
\end{array}
\right\|
\end{align}

The expectation value of the original density matrix $\rho$ \ in such a state
is $<\beta_{\pm}|\rho|\beta_{\pm}>=(1\pm z)/2.$ \ Under integration, terms
linear in $z$ vanish. \ The resulting density matrix is then:
\begin{align}
&  \frac{1}{N}\int d\widehat{n}\left\{
\begin{array}
[c]{c}%
\frac{1+z}{2}\left\|
\begin{array}
[c]{cc}%
\frac{1+z}{2} & \\
& \frac{1-z}{2}%
\end{array}
\right\|  +\\
\frac{1-z}{2}\left\|
\begin{array}
[c]{cc}%
\frac{1-z}{2} & \\
& \frac{1+z}{2}%
\end{array}
\right\|
\end{array}
\right\} \\
&  =\frac{1}{4N}\int d\widehat{n}\left\|
\begin{array}
[c]{cc}%
2(1+z^{2}) & \\
& 2(1-z^{2})
\end{array}
\right\| \\
&  =\frac{1}{4}\left\|
\begin{array}
[c]{cc}%
2(4/3) & \\
& 2(2/3)
\end{array}
\right\|  =\frac{1}{4}\left\|
\begin{array}
[c]{cc}%
8/3 & \\
& 4/3
\end{array}
\right\|
\end{align}

We see that measurement along an axis chosen at random does not mix the
original density matrix as thoroughly as does a quantum mechanical rotation
about an axis chosen at random. \ The best that can be done is to reduce the
state from one which is ``pure'', which we might represent as
$(100\%\,|+>,0\%|->)$ \ to a mixture $(2/3|+>,1/3|->).$ \ This is, however,
better thought of as being a mixture of polarized ($\rho_{p\text{ }}$) and
unpolarized ($\rho_{u}$)states.
\begin{align}
\frac{1}{3}\left\|
\begin{array}
[c]{cc}%
2 & \\
& 1
\end{array}
\right\|   &  =\frac{1}{3}\left\|
\begin{array}
[c]{cc}%
1 & \\
& 0
\end{array}
\right\|  +\frac{2}{3}\left\|
\begin{array}
[c]{cc}%
\frac{1}{2} & \\
& \frac{1}{2}%
\end{array}
\right\| \\
&  =\frac{2}{3}\rho_{u}+\frac{1}{3}\rho_{p}%
\end{align}

Clearly, this state gives Q an advantage, since the odds are 2:1 in favor of
the state in which he left the system. \ It is a little surprising that P
cannot fully depolarize the system, and I believe it is again due to the fact
that a direction chosen at random has a non-zero component along the direction
of original polarization, and is therefore unable to completely depolarize the
system. \ Note that if player P is permitted to iterate this depolarizing
measurement, the unpolarized component is unchanged by the measurement , while
the polarized component is gradually depolarized. \ After $n$ such
measurements the mixture become:
\begin{equation}
3^{-n}\rho_{p}+(1-3^{-n})\rho_{u}%
\end{equation}

To sum up, the puzzling result given in [1] was attributed to the limited
``classical nature'' of the second player. \ We find, instead, that it is due
to the fact that the first player knows exactly what the second player will
do. \ That advantage is eliminated if the second player uses a genuinely mixed
strategy, involving an axis chosen at random. \ In a real world there is in
fact no way to know which axis is aligned with the coordinates chosen by the
first player. Our more careful analysis shows that a second player with access
to unitary transformations can bring the system to one in which the expected
value of the spin operator is the same as it is in a completely unpolarized
state by a ``single'' rotation of 120%
${{}^o}$%
about a random axis. \ On the other hand, a player who is forced to reduce the
wave function each time that he touches the system, cannot depolarize the
system in a single step. \ This seems to be due to the fact that he cannot
know the orientation of the system, and to some extent will be measuring it
along the axis of polarization, which has no depolarizing effect. \ However,
if iterated measurements are permitted, about independently random axes, the
system will become as close to unpolarized as one likes. We can summarize the
possibilities in a table of Odds favoring each player (ties are ignored)..%

\begin{tabular}
[c]{lll}%
Case & Player P & Odds Q:P\\
\lbrack1] & Rotate or leave as is & 1:0\\
\lbrack2] & Rotate $120^{\circ};$ random axis & 1:1\\
\lbrack3] & Measure; random axis & 2:1
\end{tabular}

While it is hoped that this note clarifies the underlying physics of the
puzzling result given in [1], its implications for quantum computing, which is
a topic of great interest, are interesting as well. \ It suggests that
attempts to develop quantum mechanical analyses of games or other noisy
situations, for better understanding quantum computing and steganography must
be done with care to maintain a correctly quantum mechanical description
throughout. Switching, at any point, into a comparatively incomplete
''classical'' formulation can lead to paradoxical and incorrect results.

Note that while this note owes its existence to the stimulating contribution
of [1], the concepts needed for its development are given in work by John von
Neumann [1,2] before the midpoint of the last century. The author acknowledges
support of the Fulbright Foundation, for a Research Fellowship to Oslo
University College, in Norway. \ This research is supported in part by the
National Science Foundation under Grant IIS 98-12086, and by the Defense
Advanced Research Projects Agency (DARPA) under contract \ \ N66001-97-C-8537.

\section{References}

\bigskip

[1] Meyer, David A. Quantum Strategies. Physical Review Letters 82(5)1052-1055.

[2] von Neumann, John. Mathematical foundations of quantum mechanics.
translated from the German edition by Robert T. Beyer. Princeton University
Press, 1996. Princeton, N.J. \ 445 p.

[3] von Neumann, John, Morgenstern, Oskar . \ Theory of games and economic
behavior. [3d ed.], Princeton University Press, 1953 [c1944]. Princeton
\end{document}